\documentclass[journal=jctcce,manuscript=article,layout=twocolumn]{achemso}
\usepackage{graphicx}
\usepackage[version=3]{mhchem}

\author{J. R. Trail}\email{jrt32@cam.ac.uk}
\affiliation{Theory of Condensed Matter Group, Cavendish Laboratory, J
  J Thomson Avenue, Cambridge CB3 0HE, United Kingdom}

\author{R. J. Needs}
\affiliation{Theory of Condensed Matter Group, Cavendish Laboratory, J
  J Thomson Avenue, Cambridge CB3 0HE, United Kingdom}

\date{\today}

\title{Comparison of Smooth Hartree-Fock Pseudopotentials}

\begin{document}

\begin{abstract}
  The accuracy of two widely used scalar-relativistic Hartree-Fock
  pseudopotentials, the Trail-Needs-Dirac-Fock (TNDF) and the
  Burkatzki-Filippi-Dolg (BFD) pseudopotentials, is assessed.  The
  performance of the pseudopotentials is tested for a chemically
  representative set of 34 first-row molecules.  All comparisons are
  made at the Hartree-Fock level of theory, and both sets of 
  pseudopotentials give good results.  The all-electron equilibrium 
  geometries, molecular dissociation energies, and zero-point 
  vibrational energies are reproduced a little more accurately by 
  the TNDF pseudopotentials than the BFD ones.
\end{abstract}


\maketitle

\section{Introduction}

Pseudopotentials are used to improve the efficiency of electronic
structure calculations by replacing the inert core electrons by an
angular-momentum-dependent effective core potential.  This approach is
particularly advantageous for heavy atoms, but it is also useful for
light atoms including hydrogen.  The use of pseudopotentials is well 
established within density-functional-theory (DFT) and
Hartree-Fock (HF) theory, and they are also used within correlated 
quantum chemical methods and in quantum Monte Carlo (QMC)
methods.
\cite{Ceperley-Alder_1980,Foulkes_QMC_review,Casino_reference,
      Lester_review_2012,Kolorenc_QMC_review_2011}
Pseudopotentials can describe the influence of the core electrons on
the valence electrons, so that an accurate description of the chemical
bonding may be obtained.
Near the core the pseudopotentials, and therefore the
accompanying orbitals, are chosen to be as smooth as possible,
given the requirement that they reproduce the low-energy electronic
states.

In earlier work we reported the construction of a `periodic table' of
smooth pseudopotentials from the Dirac-Fock theory, which we 
shall refer to as the Trail-Needs-Dirac-Fock (TNDF) pseudopotentials
\cite{Trail_2005_asymptotic,Trail_2005_pseudopotentials}.  
Another set of smooth HF-based pseudopotentials has been provided 
by Burkatzki, Filippi and Dolg (BFD) 
\cite{Burkatzki_2007,Burkatzki_2008}.
Both the TNDF and BFD pseudopotentials have been used to calculate the 
properties of a variety of molecules\cite{mol1,mol2,mol3,mol4,mol5}, 
clusters\cite{clus1,clus2,clus3}, and bulk solids\cite{bulk1,bulk2}.
Pseudopotentials derived within the HF theory are often used in QMC
calculations.
The smoothness of the pseudopotentials is particularly advantageous
for methods using a plane-wave basis, and for QMC applications in
general.
The two sets of pseudopotentials are generated using 
fundamentally different methods, and it is instructive to compare 
the accuracy with which they reproduce all-electron (AE) HF results.

Confidence in the application of the TNDF and BFD pseudopotentials, or
comparison of their accuracy, requires an assessment of how well they
reproduce the properties of a representative set of systems at the
AE-HF level of theory.  All previous assessments and comparisons have
been somewhat limited in the physical properties, computational
errors, and variety of systems considered.  Although the
pseudopotentials may be used for molecules, clusters and condensed
matter systems, it is most convenient to test their performance using
small molecules.

In this paper, we evaluate the accuracy with which nonrelativistic 
HF calculations using the TNDF and BFD pseudopotentials reproduce 
the molecular geometry, dissociation energy, and zero-point vibrational 
energy (ZPVE) provided by relativistic AE HF theory.
We limit ourselves to the HF theory as both pseudopotential types 
were constructed to reproduce AE HF results.
Similarly, we include relativistic effects in our AE calculations, as both
the TNDF and BFD pseudopotentials were constructed to reproduce such
effects.

Two comparisons of the TNDF and BFD pseudopotentials have already been
reported in the literature\cite{Fracchia_2012,Burkatzki_2007}.
However, for both comparisons, the set of molecules tested was not large 
enough to draw reliable conclusions, and basis set error was neither 
estimated nor controlled by using extrapolation methods.
In the tests reported here, we have used a larger set of molecules than
in the previous tests, and we have carefully studied and largely
eliminated the basis set errors.

The test set of molecules was obtained by taking the neutral members of the
G1\cite{G1a,G1b} set, removing all molecules that contain atoms other
than the series H--F, and adding the H$_2$, BH, B$_2$, C$_2$, and 
NO$_2$ molecules, resulting in a test set of 34 molecules.
Note that we do not consider Be$_2$ because it is not bound within HF
theory.
We include F$_2$ because restricted HF theory provides a well-defined 
equilibrium geometry for this molecule even though the
dissociation energy takes a physically unrealistic negative
value\cite{gordon_87}.

We confine our attention to the light atoms H--F, which ensures
that both the AE and pseudopotential results suffer from negligible 
basis set error.
This allows us to ascribe most of the disagreement between these results 
to errors in the pseudopotential representation of the core electrons.
Separating the basis set and pseudopotential errors is also necessary to
test the suitability of the TNDF and BFD pseudopotentials for use in
diffusion quantum Monte Carlo (DMC) calculations\cite{Foulkes_QMC_review}.
Limiting ourselves to these light atoms also ensures that relativistic
effects are small, so that the differences due to inconsistent
implementations of relativistic effects in the generation of the two
pseudopotential types is expected to be small.
The goal of our comparison of AE and pseudopotential results is to
identify which of the pseudopotential types provides the more accurate
reproduction of the AE HF results.

Atomic units are used, unless otherwise indicated, and HF refers to 
restricted open-shell Hartree-Fock.

\section{Method}

The TNDF pseudopotentials were generated via the construction of 
pseudoatoms that reproduce the valence contribution to the AE 
wave function outside of the core region.
Inversion of the HF equations then provides an effective potential
for each angular momentum channel.
The underlying AE states were generated at the Dirac-Fock level of
theory, so that the resulting pseudopotential includes the relativistic
effects present in the Dirac-Fock theory.
This approach is similar to the standard methods used in DFT calculations;
however, for the HF theory, the long range of the exchange interaction can
lead to a finite value of the pseudopotential at large distances from
the core\cite{Trail_2005_asymptotic}.
This pathological behavior was removed in the generation procedure.
Such pseudopotentials are often referred to as `shape consistent'.

The BFD pseudopotentials were generated by taking a Gaussian
parametrization of the pseudopotential and determining parameter
values that accurately reproduce the AE total energy differences
between a number of atomic states.
The underlying AE energies were generated at the Wood-Boring HF level
of theory and hence include scalar-relativistic effects\cite{wood_78}.
Such pseudopotentials are often referred to as `energy consistent'.

Both sets of pseudopotentials include relativistic effects and are 
provided as Gaussian expansions for use in standard quantum chemistry 
packages.
The BFD pseudopotentials do not include the effects of spin-orbit 
coupling, whereas the TNDF pseudopotentials do.  
To make the comparisons as fair and useful as possible, we therefore
use only the spin-averaged TNDF pseudopotentials, and we do not use
the associated spin-orbit potentials \cite{TNDF_website}.

In the following, all calculations are performed using the
\textsc{molpro}\cite{molpro} quantum chemistry package.  All-electron
results are obtained using HF theory with the second-order scalar
relativistic Douglas-Kroll-Hess Hamiltonian, and therefore, they
include scalar relativistic effects but not spin-orbit coupling.
Results for both the TNDF and BFD pseudopotentials are generated using
nonrelativistic HF theory, and therefore, they include relativistic
effects only through the representation of the core electrons by the
pseudopotentials.  We therefore neglect the very small relativistic
effects on the exchange interactions between the valence electrons.

In order to distinguish basis set error from errors due to the 
pseudopotentials themselves, it is desirable to compare results close 
to the complete basis set limit and to use the same basis sets for each 
calculation. We use Dunning basis sets of the aug-cc-pVnZ type in their 
uncontracted form. Such basis sets provide good convergence properties 
for both AE and pseudopotential calculations.
They also consistently provide lower HF energies than the basis sets
provided with the BFD pseudopotentials\cite{BFD_website} for the 
first row diatomic molecules\cite{Trail_2013_cepp}.

A comparison of results close to the complete basis set limit is also 
appropriate for investigating the accuracy of pseudopotentials for use 
in diffusion quantum Monte Carlo (DMC) calculations, for which the 
basis set error is far smaller than that found in HF or quantum chemical 
calculations.

Optimized geometries, dissociation energies, and ZPVEs are generated 
for each molecule in our test set.
We quantify the dissociation energy as the molecular well depth, $D_e$, 
which does not include the ZPVE.
Basis set convergence error for each of these quantities is estimated
by performing each calculation with two basis sets of the same type
but different sizes, indexed by $n$.

Geometry optimization is performed by direct minimization of the 
total energy of each molecule for each basis set.
To allow the comparison of errors in bond lengths with bond angles and
dihedral angles we map each angle to the arc of a circle of radius
$1.0$~\AA\ (a typical bond length for the molecules considered).
We take the optimum geometry for the largest basis set as our estimate 
of the equilibrium geometry.  We investigate the error in this estimate 
by extrapolating each geometry parameter to the full basis set limit 
using the two-point formula
\begin{equation}
\label{geom_extrp}
x(L) = a + b \exp[-1.335L],
\end{equation}
where $L$ is the highest angular momentum present in the basis set of 
index $n$ and taking the difference between the extrapolated value
and the estimated values as the estimated error.  We do not take the 
extrapolated limit as our estimate as no error estimate accompanies it 
and because extrapolation to the complete basis set limit is less 
reliable for geometries than for total energies.

Note that this formula is the same as that used by
Feller\cite{feller_10} for three-point extrapolation of optimized
geometries of hydrocarbons but with a fixed value of the exponential
parameter.
This fixed exponential parameter was obtained by considering only the
diatomic molecules in the test set.
An additional geometry optimization for each of these was performed
using the next smallest basis set, allowing three-point extrapolation
to be performed with the exponential parameter free to vary.  The
value of $L$ used in eq\ \ref{geom_extrp} is obtained by averaging the
resulting exponential parameters over all diatomic molecules in our
test set and averaging the parameters arising from both
the pseudopotential and AE results.

Well depths, $D_e$, for each molecule are obtained as the difference 
between the total energy for each molecule and the sum of the total 
energies of the component atoms, using consistent basis sets for each.
This provides a $D_e$ for each basis set and molecule.
Three estimates of the complete basis set limit are provided by the 
three two-point extrapolation formulas\cite{jensen_05,karton_06},
\begin{eqnarray}
D_{e,1}(L)&=&a_1+b_1L^{-8.74}                         \nonumber \\
D_{e,2}(L)&=&a_2+b_2\exp[-1.95L]                      \nonumber \\
D_{e,3}(L)&=&a_3+b_3(L+1)\exp[-9\sqrt{L}],
\end{eqnarray}
which may be combined to provide an estimate and error for the complete 
basis set limit of\cite{feller_10}
\begin{equation}
D_e = (a_1+a_2+a_3)/3 \pm \textrm{Max} \left[ |a_i-D_{e}| \right].
\end{equation}

Harmonic ZPVEs are obtained within HF theory by diagonalization of 
the Hessian obtained from numerical energy derivatives at the optimum
geometry and summation of the contributions from each mode.

Unlike the optimized geometries and $D_e$, we do not use extrapolation
to estimate errors in the ZPVEs because a justification for such an
approach is not available in the literature and because the errors in
the ZPVEs calculated with a finite basis are small.

We do, however, estimate the basis set error.  The estimated 
ZPVE is taken to be that resulting from the largest basis 
set used, and the estimated error is taken to be the 
difference between the ZPVEs resulting from the two basis sets.

Due to finite computational resources and the availability of basis
sets, the choice of basis sets is somewhat complex.  For all diatomic
molecules, we use $n=5,6$, except for those containing Li or Be, for
which we use $n=4,5$.  For all other molecules, we use $n=4,5$, with
the exception of H$_2$O$_2$, H$_3$COH, H$_4$N$_2$, and C$_2$H$_6$, for
which we use $n=3,4$.  The same pair of basis sets are used for
obtaining optimized geometries, well depths, and ZPVEs for all
molecules with the exception of C$_2$H$_4$, for which we use $n=4,5$
for geometry optimizations and well depths and $n=3,4$ for ZPVEs.

Since the publication of the original paper of Burkatzki
\emph{et al.},\cite{Burkatzki_2007} a significantly improved BFD 
pseudopotential for hydrogen has been provided by Filippi and Dolg 
and used by Petruzielo \emph{et al.}\cite{Petruzielo_12}.
In the following, we present results for both the original set of BFD 
pseudopotentials, (BFD(2007)), and for the improved H pseudopotential 
(BFD(2012)). Throughout the text, all discussion and results 
associated with `BFD pseudopotentials' refers to BFD(2012) unless 
stated otherwise.

\section{Results}

First, we consider the basis set errors for our estimated geometries,
well depths, and ZPVEs.
We found all of these errors to be well within chemical accuracy of
$0.01$~\AA, $0.57^\circ$, and $1$~kcal$~$mol$^{-1}$ with the basis
set error (averaged over the AE and pseudopotential results and over 
the full set of molecules) taking the values $0.0001$~\AA,
$0.0001$~kcal$~$mol$^{-1}$, and $0.004$~kcal$~$mol$^{-1}$ for geometry
parameters $D_e$ and ZPVEs, respectively.

For the well depths and ZPVEs, the basis set error is negligible, with 
peak errors less than $0.06\%$ of chemical accuracy.
For the geometry parameters, the error is acceptably small for all 
molecules, at a few percent of chemical accuracy for most 
and with a maximum of $15\%$ of chemical accuracy for the dihedral 
angle of H$_2$O$_2$.

Our goal is to assess the accuracy with which the pseudopotentials
reproduce the AE results; hence, we take our AE results as the baseline 
and consider only the deviation of the pseudopotential results from 
this baseline.
Because this is a difference between calculated quantities,
basis set error is further reduced by correlations between the errors
present in the AE and pseudopotential results.

The basis set error in the difference between our pseudopotential and
AE well depths and ZPVEs remains negligible, and the error in the 
differences for geometry parameters is not significant with a mean 
basis set error of $0.0001$~\AA\ for geometry differences.

We summarize the pseudopotential error over the test set in terms of 
the mean absolute deviation (MAD) of the pseudopotential results 
from the AE results, and the maximum absolute deviation of the 
pseudopotential results from the AE results. The data are provided in 
Table~\ref{tab:1}.

\subsection{Optimized Geometries}

\begin{figure*}[t]
\includegraphics[scale=1.00]{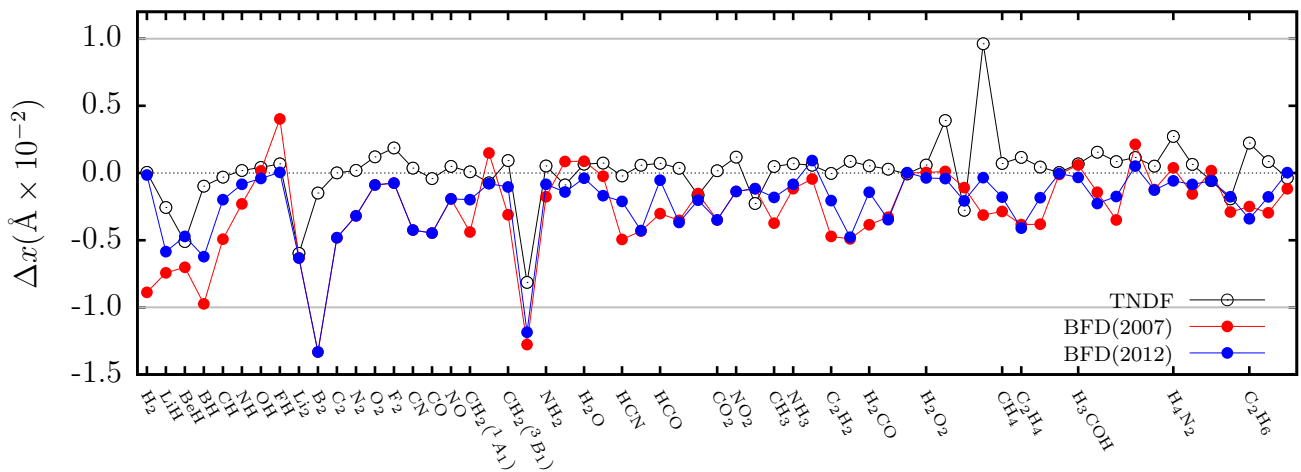}
\caption{ \label{fig1}
  Deviation of the spatial parameters from the AE HF reference data 
  for the TNDF and BFD pseudopotentials.  The horizontal gray solid 
  lines at $\pm$0.01~\AA\ indicate the upper and lower limits of 
  chemical accuracy.
}
\end{figure*}

Figure~\ref{fig1} shows the errors in the optimum geometry parameters
for the two pseudopotential types.
Overall, the TNDF pseudopotentials reproduce the AE geometries more
accurately than the BFD ones, with a few exceptions.
The MAD (with respect to the AE results) arising with the BFD 
pseudopotentials is $1.8\times$ greater than that for the TNDF 
pseudopotentials.
There is a trend for the BFD pseudopotentials to underestimate bond
lengths, with geometry parameters showing an average deviation from
the AE results of $-0.002$~\AA, whereas for the TNDF pseudopotentials
the average deviation is only $0.0001$~\AA.
Unlike the BFD pseudopotentials, all of the geometries obtained with
the TNDF pseudopotentials fall within chemical accuracy of the AE
geometries.

The geometry of H$_2$O$_2$ stands out as the worst case for the TNDF
pseudopotentials, with a significantly larger error than the
equivalent result with BFD pseudopotentials.  We have not found an
explanation for this difference, but we note that the shallow
variation of the total energy with geometry for this molecule is known
to result in an optimum geometry that is unusually sensitive to errors
in the total energy.

\subsection{Well Depths}

\begin{figure*}[t]
\includegraphics[scale=1.00]{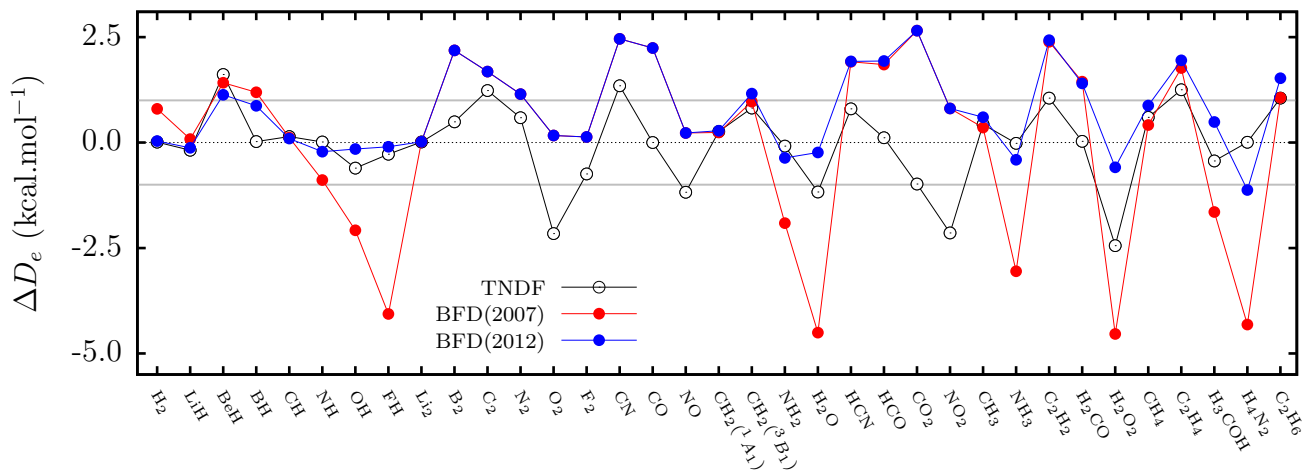}
\caption{ \label{fig2} Deviation of the well depths from the AE HF
  reference data for the TNDF and BFD pseudopotentials. The horizontal
  gray solid lines at $\pm$1~kcal$~$mol$^{-1}$ indicate the upper and
  lower limits of chemical accuracy.  The estimated basis set error is
  negligible on the scale of the figure.  
}
\end{figure*}

Figure~\ref{fig2} shows the errors in well depths at optimum geometries 
for the two pseudopotential types.
Overall the TNDF pseudopotentials reproduce the AE $D_e$ values more 
accurately than the BFD pseudopotentials, with error in well depths 
for the TNDF pseudopotentials smaller than that for the BFD 
pseudopotentials for $24$ out of the $34$ molecules studied.
In summary, the MAD (with respect to the AE results) for the BFD 
pseudopotentials is $1.4\times$ greater than that for the TNDF 
pseudopotentials.
The BFD pseudopotentials tend to overestimate $D_e$, with an average 
deviation from AE values of $0.78$~kcal$~$mol$^{-1}$, whereas for the 
TNDF pseudopotentials the average deviation is 
$-0.015$~kcal$~$mol$^{-1}$.

Figure~\ref{fig2} also shows the errors in well depth that arise with
the original BFD hydrogen pseudopotential (BFD(2007)).
The poor performance of the original hydrogen pseudopotential is most 
apparent for molecules containing H bonded to N, O, or F, the most 
electronegative atoms considered.
On examination of the updated pseudopotentials, it is clear that the 
poor transferability of the original BFD pseudopotential can be 
ascribed to it deviating from the Coulomb potential over a core region 
that is too large.

\subsection{Harmonic Zero Point Vibrational Energies}

\begin{figure*}[t]
\includegraphics[scale=1.00]{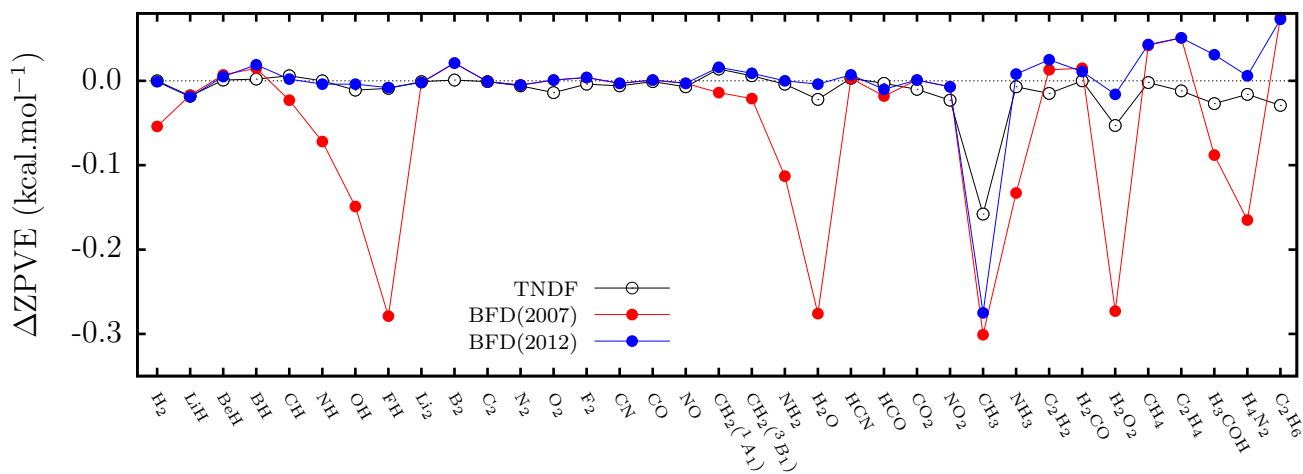}
\caption{ \label{fig3}
  Deviation of the ZPVEs from the baseline AE HF reference data 
  for the TNDF and BFD pseudopotentials.
}
\end{figure*}

Figure~\ref{fig3} shows the errors in the ZPVEs for the two 
pseudopotential types.
Overall, the TNDF pseudopotentials reproduce the AE ZPVEs more
accurately than the BFD pseudopotentials; the MAD (with respect to 
the AE results) for the BFD pseudopotentials is $1.4\times$ greater 
than that for the TNDF pseudopotentials.
However, the errors themselves are considerably smaller than chemical 
accuracy and both pseudopotential types can be considered as
chemically accurate for ZPVEs.

Figure~\ref{fig3} also clearly shows that for the original BFD hydrogen 
pseudopotential (BFD(2007)) the errors in the ZPVEs are dominated by the 
poor description of H bonding with N, O, or F.
The data also demonstrates that the updated BFD hydrogen pseudopotential 
removes most of this error, with a large reduction in the error 
associated with a few molecules reducing the BFD MAD by a factor of 
$0.31\times$.

\begin{table*}[b]
\begin{tabular}{rrrrrrr}									\\ \hline \hline 
     & \multicolumn{2}{c} { Geometry~(\AA) }
     & \multicolumn{2}{c} { $D_e$~(kcal$~$mol$^{-1}$) }
     & \multicolumn{2}{c} { ZPVE~(kcal$~$mol$^{-1}$) } \vspace{-6pt}	\\
     & \multicolumn{2}{c} \hrulefill 
     & \multicolumn{2}{c} \hrulefill 
     & \multicolumn{2}{c} \hrulefill \\
     & \multicolumn{1}{r}{MAD}
     & \multicolumn{1}{r}{Max}
     & \multicolumn{1}{r}{MAD}
     & \multicolumn{1}{r}{Max}
     & \multicolumn{1}{r}{MAD}
     & \multicolumn{1}{r}{Max}       \\ \hline
\multicolumn{1}{l}{TNDF}      & 0.00130 &  0.00962 & 0.69457 & -2.44325 & 0.01406 & -0.15800 \\
\multicolumn{1}{l}{BFD(2012)} & 0.00230 & -0.01332 & 0.96420 &  2.65210 & 0.01989 & -0.27500 \\ 
\multicolumn{1}{l}{BFD(2007)} & 0.00313 & -0.01332 & 1.62093 & -4.53708 & 0.06474 & -0.30100 \\ \hline \hline
\end{tabular}
\caption{ \label{tab:1}
  Deviation of Pseudopotential Results from AE Results.
  Pseudopotential errors are summarized in terms of the mean absolute deviation (MAD) of 
  the pseudopotential results from the AE results, and as the maximum deviation (Max) of 
  the pseudopotential results from the AE results (over the test set of molecules).
  Considering only the TNDF and BFD(2012) data sets, the maximum error for geometry 
  parameters, dissociation energies, and ZPVEs occurs for the B$_2$, CO$_2$, and CH$_3$ 
  molecules described by the BFD(2012) pseudopotentials.
}
\end{table*}

\section{Conclusions}

Due to the large test set used, our comparison of the performance 
of the TNDF and BFD pseudopotentials at the HF level of theory
is more reliable than those of previous authors.
Because the pseudopotentials themselves are constructed to reproduce AE
HF results, performing our analysis using the HF theory does not limit the
validity of our conclusions.
We believe that the accuracy comparison provided here is more complete
than those of previous authors \cite{Fracchia_2012,Burkatzki_2007}.
We have explicitly controlled the basis set error, have considered
optimum geometries and ZPVEs as well as dissociation energies, and
have considered a larger set of molecules chosen to be representative
of the chemical properties of the first row atoms.

For the large test set of molecules considered, the TNDF 
pseudopotentials reproduce (relativistic) AE results 
for optimized geometries, well depths and ZPVEs to a modestly
higher accuracy than the BFD pseudopotentials.
The MADs of the pseudopotentials results from the AE results are
$1.8\times$, $1.4\times$, and $1.4\times$ greater for the BFD
pseudopotentials than for the TNDF pseudopotentials for geometry
parameters, well depths, and ZPVEs, respectively.
The main limitation of this study is that only first row atoms have
been considered, and it would be premature to draw conclusions about
the performance of the two pseudopotential types for heavier atoms.

Overall, the small improvement of the TNDF over the BFD
pseudopotentials is most significant for optimized geometries and
dissociation energies.  The calculated ZPVEs are well within chemical
accuracy.

\begin{acknowledgement}
  The authors were supported by the Engineering and Physical Sciences
  Research Council (EPSRC) of the United Kingdom.
\end{acknowledgement}


\end{document}